\documentclass[twocolumn]{aastex631}
\usepackage{physics}
\usepackage{gensymb}
\usepackage{CJK}

\DeclareMathOperator*{\diag}{diag}

\newcommand{\Eq}[1]{\text{Eq.\,\ref{#1}}}
\newcommand{\Figure}[1]{\text{Figure \ref{#1}}}

\newcommand{\vbd}{\vb{d}}
\newcommand{\vbm}{\vb{m}}

\newcommand{\vbn}{\vb{n}}
\newcommand{\vbb}{\vb{b}}
\newcommand{\inv}[1]{#1^{-1}}

\newcommand{\hatm}{\vb{\hat{m}}}
\newcommand{\Pdagger}{P^{\dagger}}
\newcommand{\Nbar}{\bar{N}}
\newcommand{\PPinv}[1]{\inv{\qty(\Pdagger #1 P)}}
\newcommand{\Neta}{N_{\eta}}

\graphicspath{{./}{figures/}}

\begin{document}
\begin{CJK*}{UTF8}{gbsn}

\title{Cooling Improves Cosmic Microwave Background Map-Making When Low-Frequency Noise is Large}

\author[0000-0002-2981-4951]{Bai-Chiang Chiang (强百强)} 
\affiliation{Department of Physics, Florida State University, Tallahassee, Florida 32306}

\author[0000-0001-7109-0099]{Kevin M. Huffenberger}
\affiliation{Department of Physics, Florida State University, Tallahassee, Florida 32306}

\begin{abstract}

In the context of Cosmic Microwave Background data analysis, we study the solution to the equation that transforms scanning data into a map.
As originally suggested in ``messenger'' methods for solving linear systems, we split the  noise covariance into uniform and non-uniform parts and adjust their relative weights during the iterative solution.
With simulations, we study mock instrumental data with different noise properties, and find that this ``cooling'' or perturbative approach is particularly effective when there is significant low-frequency noise in the timestream.  In such cases, a conjugate gradient algorithm applied to this modified system converges faster and to a higher fidelity solution than the standard conjugate gradient approach.
We give an analytic estimate for the parameter that controls how gradually the linear system should change during the course of the solution.

\end{abstract}


\section{Introduction} \label{sec:intro}

In observations of the Cosmic Microwave Background (CMB), map-making is an intermediate step between the collection of raw scanning data and the scientific analyses, such as the estimation of power spectra and cosmological parameters.
Next generation CMB observations will generate much more data than those today, and so
it is worth exploring efficient ways to process the data even though, on paper, the map-making problem has long been solved.

The time-ordered scanning data is summarized by
\begin{equation}
\vbd = P\vbm + \vbn \label{map making model}
\end{equation}
where $\vbd$, $\vbm$, and $\vbn$ are the vectors of time-ordered data (TOD), the CMB sky-map signal, and measurement noise.
$P$ is a sparse matrix in the time-by-pixel domain that encodes the telescope's pointing.
Of several map-making methods \citep{1997ApJ...480L..87T}, one of the most common is the method introduced for the Cosmic Background Explorer \cite[COBE,][]{1992ASIC..359..391J}.  This optimal, linear solution is 
\begin{align}
\qty(\Pdagger \inv{N}  P) \hatm = \Pdagger \inv{N} \vbd \label{map making eq}
\end{align}
where  $\hatm$ provides the standard generalized least squares minimization of the $\chi^2$ statistic,

\begin{align}
\chi^2(\vbm) & \equiv (\vbd - P\vbm)^{\dagger} N^{-1} (\vbd - P \vbm).
\label{chi2 formula}
\end{align}
Here we assume that the noise has zero mean $\ev{\vbn} = \vb{0}$,
and the noise covariance matrix $N = \ev{\vbn \vbn^{\dagger}}$ is diagonal in frequency space.
In the case where the noise is Gaussian, the COBE solution is also the maximum likelihood solution.

With current computational power, we cannot solve for $\hatm$
by calculating $\PPinv{\inv{N}} \Pdagger \inv{N} \vbd$ directly.
The noise covariance matrix $N$ is often sparse in the frequency domain and the pointing matrix $P$ is sparse in the time-by-pixel domain.
In experiments currently under design, there may be $\sim 10^{16}$ time samples and $\sim 10^{9}$ pixels, so these matrix inversions are intractable unless the covariance is uniform (proportional to the identity matrix $I$).
We can use iterative methods, such as conjugate gradient descent, to avoid the matrix inversions, and execute each matrix multiplication in a basis where the matrix is sparse, using a fast Fourier transform to go between the frequency and time domain.

As an alternative to conjugate gradient descent, \citet{Huffenberger_2018} showed that the ``messenger'' iterative method could be adapted to solve the linear map-making system, based on the  approach from \cite{2013A&A...549A.111E} to solve the linear Wiener filter.
This technique splits the noise covariance into a uniform part and the remainder, and introduces an additional vector that represent the signal plus uniform noise.
This messenger field acts as an intermediary between the signal and the data and has a covariance that is conveniently sparse in every basis.
\cite{2013A&A...549A.111E} also introduced a cooling scheme that takes advantage of the split covariance: over the course of the iterative solution, we adjust the relative weight of those two parts.  Starting with the uniform covariance, the modified linear system gradually transforms to the final system, under the control of a cooling parameter.  In numerical experiments, \citet{Huffenberger_2018} found that a map produced by the cooled messenger method converged significantly faster than for standard conjugate gradient methods, and to higher fidelity, especially on large scales.  

\citet{2018A&A...620A..59P} showed that the the messenger field approach is equivalent to a fixed point iteration scheme, and studied its convergence properties in detail.  Furthermore, they showed that the split covariance and the modified system that incorporates the cooling can be solved by other means, including a conjugate gradient technique, which should generally show better convergence properties than the fixed-point scheme.
With simulations, we have confirmed  this conclusion.
In their numerical tests, \citet{2018A&A...620A..59P} did not find benefits to the cooling modification of the map-making system, in contrast to the findings of \citet{Huffenberger_2018}.

In this paper, we show that the difference arose because the numerical tests in \citet{2018A&A...620A..59P} used much less low-frequency (or $1/f$) noise than \citet{Huffenberger_2018}, and show that the cooling technique improves map-making performance especially when the low-frequency noise is large.  This performance boost depends on a proper choice for the pace of cooling.  \citet{2017MNRAS.468.1782K} showed that for Wiener filter the cooling parameter should be chosen as a geometric series.  In this work, we give an alternative interpretation of the parameterizing process and show that for map-making the optimal choice (unsurprisingly) is also a geometric series.

In Section \ref{sec:methods} we describe our methods for treating the map-making equation and our numerical experiments.
In Section \ref{sec:results} we present our results. 
In Section \ref{sec:conclusions}, we list our conclusions.
In Appendix \ref{appendix:eta calculation} we derive the prescription for our cooling schedule.

\section{Methods}\label{sec:methods}

\subsection{Parameterized Conjugate Gradient Method}
The messenger field approach introduced an extra cooling parameter $\lambda$ to the
map-making equation, and solved the linear system with an alternative parameterized covariance $N(\lambda) =  \lambda \tau I + \Nbar $.
The parameter $\tau = \min(\diag(N))$ represents the uniform level of (white) noise in the original covariance.
The remainder $\Nbar \equiv N - \tau I$ is the non-uniform part of the original noise covariance.
(Here $N$ without any arguments denotes the original noise covariance matrix $N = \ev{\vbn \vbn^{\dagger}}$.)
In this work we find it more convenient to work with the reciprocal of cooling parameter $\eta = \lambda^{-1}$
which represents the degree of heteroscedasticity (non-uniformity) in the parameterized covariance
\begin{equation}
  N(\eta) = \tau I +  \eta \Nbar.
\end{equation}
When $\eta=1$ this parameterized covariance $N(\eta)$ equals $N$.

\citet{2018A&A...620A..59P} showed that the conjugate gradient method can be easily applied to the cooled map-making problem.
In our notation, this is equivalent to iterating on the parameterized map-making equation
\begin{align}
  \qty(\Pdagger \inv{N(\eta_i)} P)\, \hatm(\eta_i) = \Pdagger \inv{N(\eta_i)} \vbd,
\label{map making eq eta}
\end{align}
as we adjust the parameter through a set of levels $\{\eta_i\}$.
(We use $\hatm$ with no $\eta$ argument to
mean the estimated $\hatm$ in \Eq{map making eq},
independent of $\eta$.)
This equation leads to the same system as the paramaterized equation in the messenger field method,
because $N(\eta) = \lambda^{-1} N(\lambda)$ and the condition number do not change upon scalar multiplication to both sides of the equation.
For concreteness we fix the preconditioner to $M= \Pdagger P$ for all calculations.

When $\eta = 0$, the noise covariance matrix $N(0)$ is homoscedastic (uniform), and the solution is given by the simple binned map
$\hatm(0) = \inv{\qty(\Pdagger P)} \Pdagger \vbd$,
which can be solved directly.

Since the non-white part $\bar N$ is the troublesome portion of the covariance, we can think of the $\eta$ parameter as increasing the heteroscedasticity of the system,
adding a perturbation to the solution achieved at a particular stage,
building ultimately upon the initial uniform covariance model.
Therefore, this quasi-static process requires $\eta$ increase as $0 = \eta_0 \leq \eta_1 \leq \cdots \leq \eta_{\rm final} = 1$, at which point we arrive at the desired map-making equation,
and the solution $\hatm(1) = \hatm$.

We may iterate more than once at each intermediate $\eta_i$:
we solve equation (\ref{map making eq eta}) with conjugate gradient iterations
using the result from the previous calculation $\hatm(\eta_{i-1})$ as the initial value.
We move to next parameter $\eta_{i+1}$ when the norm of residual vector 

\begin{align}
\norm{\vb{r}(\vbm,\eta_i)} \equiv
\norm{ \Pdagger \inv{N(\eta_i)} P\, 
\vbm - \Pdagger \inv{N(\eta_i)} \vbd  }
\end{align}
is an order of magnitude smaller than the norm of the right hand side of \Eq{map making eq eta}.
\begin{align}
\norm{\vb{r}(\vbm,\eta_i)} < 0.1\ \norm{\Pdagger N(\eta_i)^{-1} \vbd}
\label{norm threshold}
\end{align}
This is not stringent enough to completely converge at this $\eta$-level, but we find that it causes the system to converge sufficiently to allow us to move on to the next $\eta$.

\subsection{Analytical expression for $\qty{\eta_i}$ series}
The next question is how to appropriately choose these monotonically increasing parameters
$\eta$. 
We also want to determine $\eta_1, \cdots, \eta_{n-1}$ before starting conjugate
gradient iterations,
because the time ordered data $\vbd$ is very large,
and we do not want to keep it in the system memory during calculation or repeatedly read it in from disk.
If we determine $\eta_1, \cdots, \eta_{n-1}$ before the iterations, 
then we can precompute the right-hand side of \Eq{map making eq eta}
for each $\eta_i$ and keep these map-sized objects, instead of the entire time-ordered data.

In Appendix~\ref{appendix:eta calculation}, we show that a generic good choice for the $\eta$ parameters is {given by this} geometric series
\begin{align}
\eta_i =\min \qty\bigg{ \qty(2^i -1)\frac{\tau}{\max(\Nbar_f)},\; 1 },
\label{etai rule}
\end{align}
where $\bar N_f$ are the eigenvalues of $\Nbar$ under frequency representation.
This is one of our main results.
It {not only tells us} how to choose parameters $\eta_i$,
but also when we should stop the perturbation, and set $\eta = 1$.
For example, if {the} noise covariance matrix $N$ is almost {uniform},
then $\Nbar = N - \tau I \approx 0$,
and we would have ${\tau}/{\max(\Nbar_f)} > 1$.
This tell us that we don't need to use the parameterized method at all, 
because $\eta_0=0$ and $\eta_1= \eta_2 = \cdots= 1$.
This corresponds to the standard conjugate gradient method with simple binned 
map as the initial guess (as recommended by \citealt{2018A&A...620A..59P}).

\subsection{Intuitive Interpretation of $\eta$}\label{intuitive interp}

Here is a way to interpret the role of $\eta$ that is less technical than Appendix \ref{appendix:eta calculation}.
Our ultimate goal is to find $\hatm(1)$ which minimizes $\chi^2(\vbm)$ in \Eq{chi2 formula}.
Since $N$ is diagonal in frequency space,
$\chi^2$ could be written as a sum of all frequency {modes}
$\qty|(\vbd-P\vbm)_f|^2$ with weight $\inv{N}_f$, such as
$\chi^2(\vbm) = \sum_f \qty|(\vbd-P\vbm)_f|^2 \inv{N}_f$.
The weight is large for low-noise frequency modes (small $N_f$), and small for high-noise modes.
Which means $\chi^2(\vbm)$ would favor the low-noise modes, and therefore {the} conjugate gradient map-making focuses on minimizing the error
$\vb*{\varepsilon} \equiv \vbd - P\vbm$ in the low-noise part.

After introducing $\eta$, we minimize
$\chi^2(\vbm,\eta)$ in \Eq{chi2 eta formula} instead.
For $\eta=0$, $N^{-1}(0) \propto I$ the system is homoscedastic and the estimated map $\hatm(0)$
does not prioritize any frequency {modes}.
As we slowly increase $\eta$, we decrease the weight for the high-noise modes,
and focusing minimizing error for {the} low-noise part.
If we start with $\eta_1=1$ directly, which corresponds to the vanilla conjugate
gradient method, then {the algorithm}
will focus most on minimizing the low-noise part, such that $\chi^2$ would
converge very fast on {the} low-noise modes (typically high temporal frequencies and small spatial scales), but slowly on the high-noise modes (low frequencies and large scales).
However by introducing the $\eta$ parameter, we let the solver first treat every
frequency equally.
Then as $\eta$ slowly increases, it gradually {gives} more focus to the lowest noise part.

\subsection{Computational Cost}
To properly compare the performance cost of this method with respect to {the} vanilla
conjugate gradient method with {the} simple preconditioner,
we need to compare their computational cost at each iteration.
{
We could define $A(\eta) \equiv \Pdagger \inv{N(\eta)} P$ and $\vbb(\eta) \equiv \Pdagger \inv{N(\eta)} \vbd$,
and equation \ref{map making eq eta} could be written as $A(\eta_i) \hatm(\eta_i) = \vbb(\eta_i)$.  The right-hand side
$\vbb(\eta_i)$ could be computed before iterating,
}
since we have determined $\qty{\eta_i}$ in advance,
so it will not introduce extra computational cost.
The most demanding part of conjugate gradient method is calculating
its left hand side $A(\eta_i) \vbm$, because it contains a Fourier transform of
$P\vbm$ from the time domain to frequency domain and an inverse Fourier transform
of $\inv{N(\eta_i)} P \vbm$ from the frequency domain back to time domain,
which is order $\mathcal{O}(n\log n)$ with $n$ being the length of time ordered
data.
Compared to the traditional conjugate gradient method,
we swap $\inv{N}$ with $\inv{N(\eta)}$, and the cost is the same for one step,
since both methods need a fast Fourier transform and inverse fast Fourier transform 
at one iteration.

At each $\eta_i$ level, we use the residual to determine whether to switch to the next level ($\eta_{i+1}$), as  is Equation (\ref{norm threshold}).
Calculation of the residual vector $\vb{r}(\vbm,\eta_i)$ is part of the conjugate gradient algorithm,
so this will not add extra cost either.
Therefore, overall introducing the $\eta$ will not have extra computational cost within the conjugate gradient iterations.

However, we start a new conjugate gradient algorithm whenever $\eta_i$ updates to $\eta_{i+1}$.
Thus we must re-initialize the conjugate gradient algorithm, re-calculating the residual $\vb{r}(\vbm, \eta_{i+1})$ based on new $\eta_{i+1}$.  
This residual calculation contains an extra $A(\eta_i) \vbm$ operation.
Therefore, if we have a series $\eta_1, \eta_2, \eta_3, \cdots, \eta_{n_{\eta}}$,
there will have $n_{\eta}-1$ extra $A(\eta) \vbm$ operations compare to {the} traditional conjugate gradient
method.
If the total number of iterations is much larger than $n_{\eta}$,
then this extra cost is negligible.
For our simulation, this extra step would have rather significant impact on final result.
To have a fair comparison between the parameterized and traditional conjugate gradient method,
we will present our results with number of $\Pdagger \inv{N(\eta)} P\vbm$ operations as horizontal axis.

\subsection{Numerical Simulations}

\begin{figure}[tb!]
\includegraphics[width=\linewidth]{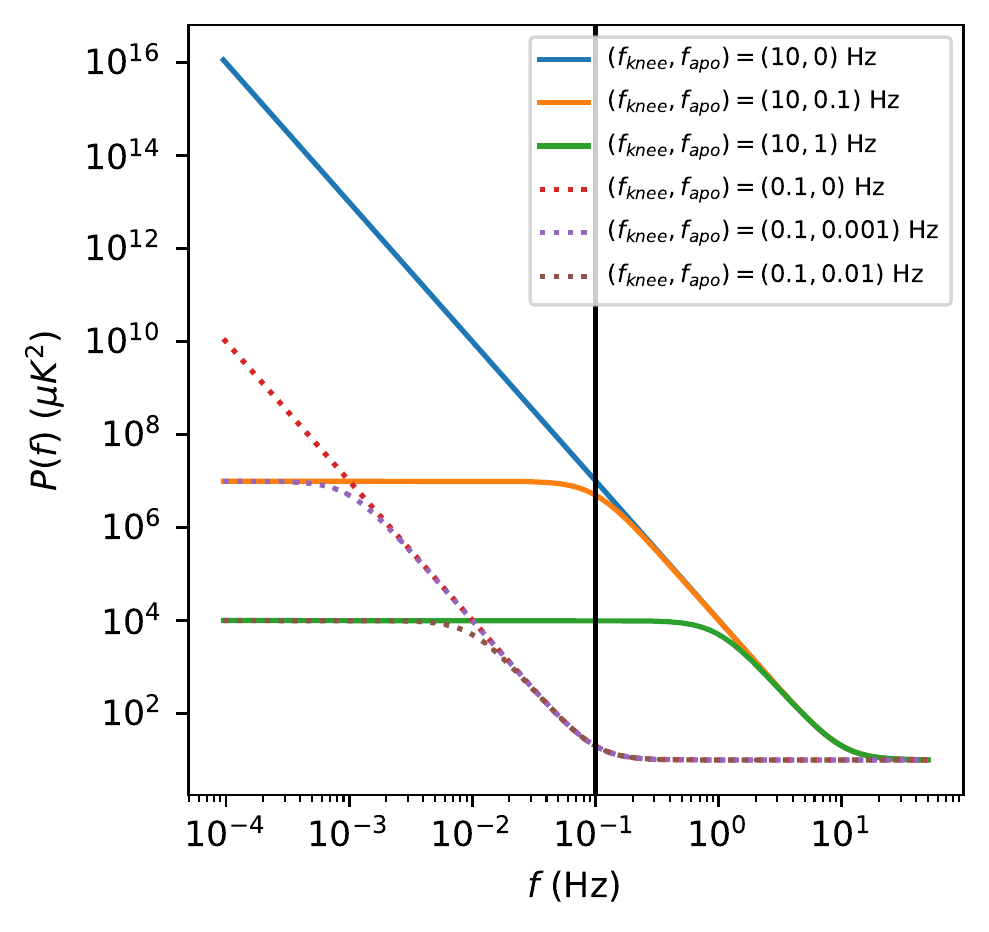}
\centering
\caption{Noise power spectra that we use in our map-making simulations.
    These show a variety of low-frequency behavior,
    parameterized by \Eq{noise power spectrum},
    with white noise at high frequency and a low-frequency power-law slope $\alpha = 3$.
    Here we show two knee frequencies, $f_\text{knee}=10$ Hz (solid lines)  
    and $f_\text{knee}=0.1$ Hz (dashed lines).
    For each knee frequency, we have shown an unflattened spectrum ($f_\text{apo}=0$ Hz), and two flattened ones ($f_\text{apo}=0.1f_\text{knee}$ and
    $0.01f_\text{knee}$).
    The vertical line shows our scanning frequency.
}
\label{power spectrum}
\end{figure}

To compare these algorithms, we need to do some simple {simulations} of scanning
processes, and generate {the} time ordered data from a random sky signal.\footnote{
The source code and other information are available at \url{https://github.com/Bai-Chiang/CMB_map_making_with_cooling}
}
Our sky is a small rectangular area, with two orthogonal directions $x$ and
$y$, both with range from $-1\degree$ to $+1\degree$.
The signal has Stokes parameters $(I,Q,U)$ for intensity and linear polarization.

For the scanning process, our mock telescope contains nine detectors,
each with different sensitivity to polarization $Q$ and $U$.
It scans the sky with a raster scanning pattern.
Its scanning frequency is $f_{\text{scan}} = 0.1$ Hz and sampling frequency is $f_{\text{sample}} = 100$ Hz.
The telescope scans the sky horizontally then vertically.
This gives the noiseless signal $\vb{s}$.
The sky signal  in the timestream has a root-mean-square (RMS) of $56$ $\mu$K.
The signal is continuous, so that it has structure on sub-pixel scales, but we find that our main conclusions remain the same when the input signal is pixelized.
In map-making, we digitize the position $(x, y)$ into $512\times 512$ pixels.

We model the noise power spectrum with
\begin{align}
P(f) = \sigma^2 \qty(1+ \frac{f_{\text{knee}}^{\alpha}+f_{\text{apo}}^{\alpha}}
    {f^{\alpha}+f_{\text{apo}}^{\alpha}}) \label{noise power spectrum}
\end{align}
which is white at high frequencies, a power law below the knee frequency, and gives us the option to flatten the low-frequency noise below an apodization frequency \citep[like in][]{2018A&A...620A..59P}.
Note that as $f_{\text{apo}} \rightarrow 0 $,
$P(f) \rightarrow \sigma^2\qty(1 + (f/f_{\text{knee}})^{-\alpha} )$, 
and it becomes a $1/f$-type noise model.

\citet{2013ApJ...762...10D} measured the slopes of the atmospheric noise in the Atacama under different water vapor conditions, finding $\alpha = 2.7$ to $2.9$.
Here we use $\sigma^2 = 10$ $\mu$K$^2$, $\alpha=3$, and compare the performance under different noise
models.  In our calculations, we choose different combinations of $f_\text{knee}$ and $f_\text{apo}$ as in \Figure{power spectrum}.  The noise spectra with the most low frequency noise have high $f_\text{knee}$ or low cut-off $f_{\text{apo}}$.

The noise covariance matrix 
\begin{equation}
N_{ff'} = P(f) \frac{\delta_{ff'}}{\Delta_f}
\label{noise covariance matrix}
\end{equation}
is a diagonal matrix in frequency space, where $\Delta_f$ is equal to {the} reciprocal
of total scanning time $T \approx 1.05\times 10^{4}$ seconds.

Finally, we get the simulated time ordered data $\vb{d} = \vb{s} + \vb{n}$ by
adding up {the} signal and noise.

\section{Results} \label{sec:results}

\begin{figure*}[tb!]
\centering
\includegraphics[width=\textwidth]{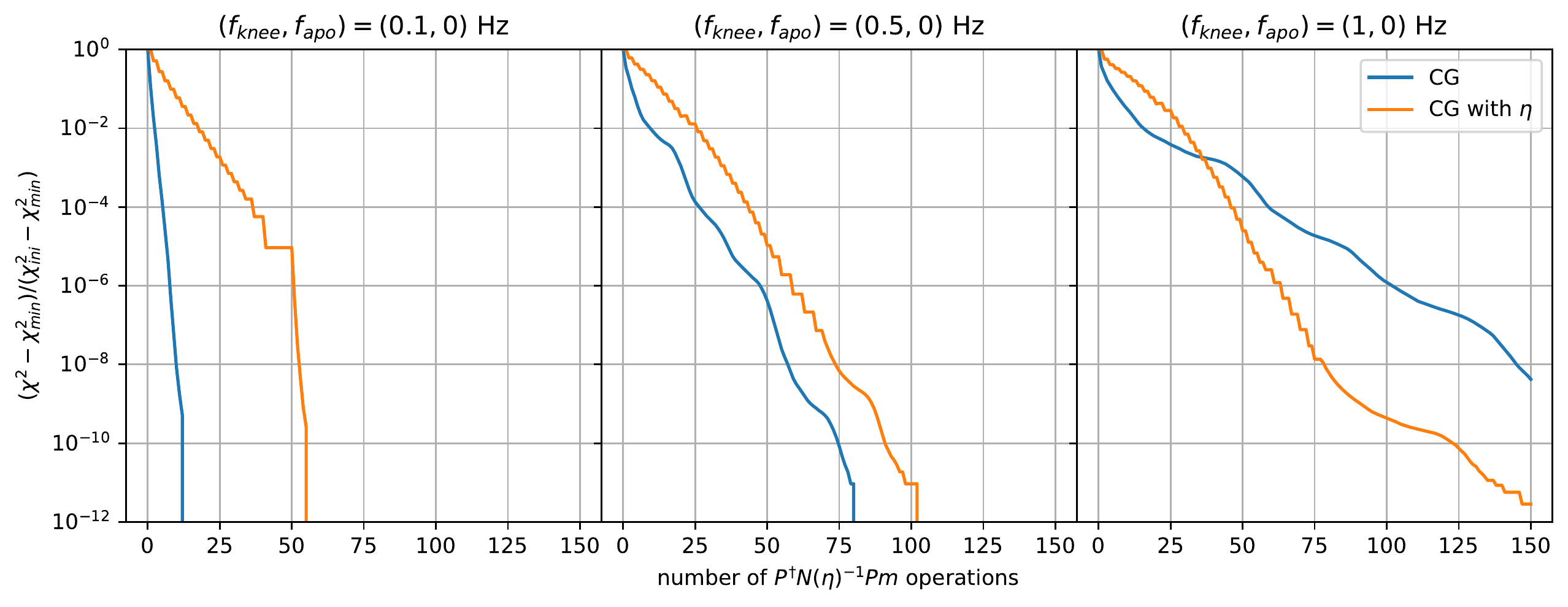}
\caption{
    Convergence properties depend on the amount of low-frequency noise,
    which increases from the left panel to the right panel with increasing knee frequency.
    The map-making equation \ref{map making eq} minimize the $\chi^2(\vbm)$, so
    the curve which falls fastest versus the number of operations is the preferred method.
    We compare the traditional conjugate gradient method (``{CG},'' blue line)
    with the parameterized conjugate gradient method (``{CG with $\eta$},'' orange line)
    under different $1/f$ noise models (fixed $f_\text{apo}=0$ Hz but different $f_\text{knee}$ in \Eq{noise power spectrum}).
    When $f_\text{knee} \gtrsim 10\,f_\text{scan} = 1$ Hz, the significant amount of
    low-frequency noise causes the parameterized conjugate gradient method to start showing its advantage.  The vertical axis is rescaled  such that all curves start from 1.
}
\label{1/f noise chi2}
\end{figure*}

\begin{figure*}[tb!]
\centering
\includegraphics[width=\textwidth]{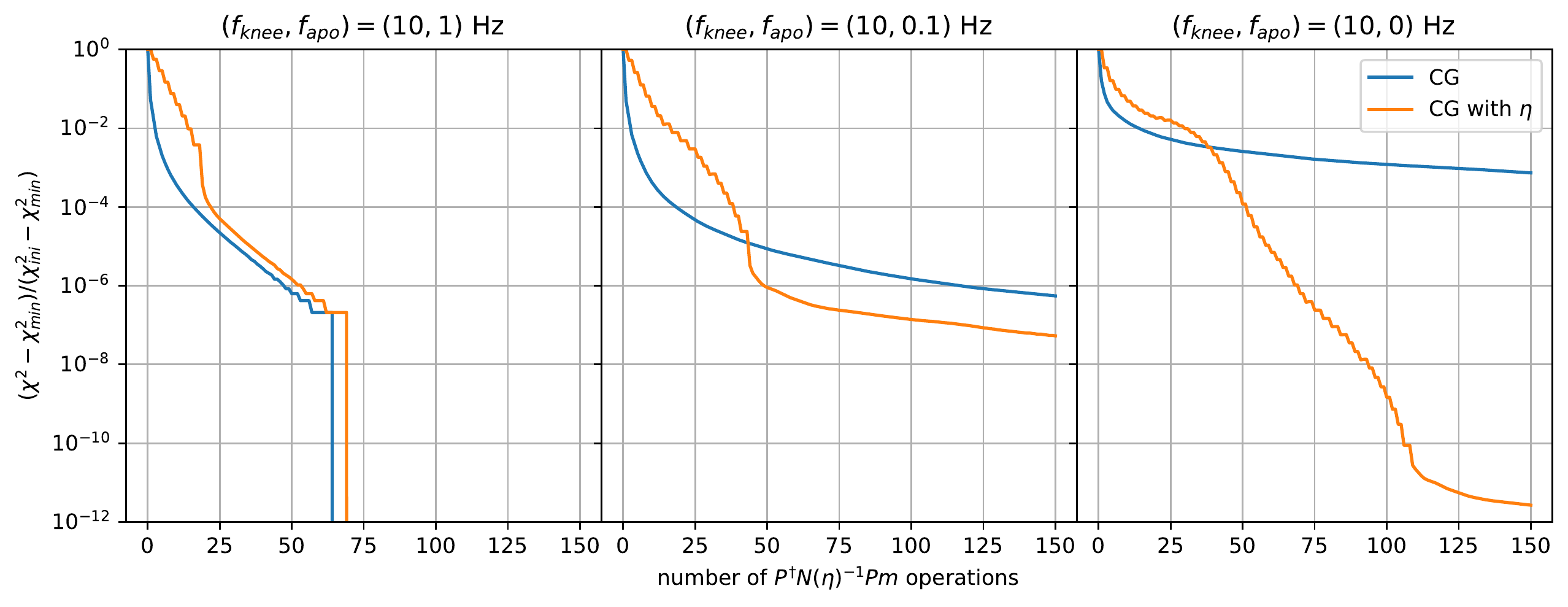}
\caption{Like \Figure{1/f noise chi2}, low-frequency noise increases from left to right, but by flattening the low-frequency noise at an apodization frequency.
    Low-frequency noise increases with decreasing apodization frequency (compare Figure \ref{power spectrum}).
    We again compare the traditional conjugate gradient method (``{CG},'' blue line) 
    with the parameterized conjugate gradient method (``{CG with $\eta$},'' orange line).
    When $f_\text{apo}$ is much smaller than $f_\text{knee}$, there is a lot of low-frequency noise and
    the parameterized conjugate gradient method is better (ultimately falls faster) than the traditional one.
}
\label{apo noise chi2}
\end{figure*}

{
We compare the standard conjugate gradient method versus the conjugate gradient with our perturbed linear system.
Both methods use the simple preconditioner $\Pdagger P$.
}
\Figure{1/f noise chi2} shows {the $\chi^2$ results} for {$1/f$ noise models} ($f_\text{apo}=0$)  with different knee frequencies.
Note that the $\chi^2$ values in all figures are calculated based on the standard $\chi^2(\vbm)$ in \Eq{chi2 formula},
not the $\eta$-dependent $\chi^2(\vbm, \eta)$ of the modified system (\Eq{chi2 eta formula}).
The minimum $\chi^2_{\text{min}}$ that we use for comparison is calculated from a deliberately slowed and well-converged parameterized conjugate gradient
method: one with 100 $\eta$ values and that halts when the final norm of the residual $\norm{\vb{r}(\vbm,1)}$
is smaller than $10^{-5}\times\norm{\Pdagger N^{-1} \vbd}$, or 100 iterations after $\eta=1$.
From \Figure{1/f noise chi2}, we can see for the $1/f$ noise model,
when $f_\text{knee} \gtrsim 10 f_\text{scan}$ the parameterized method starts showing an advantage
over {the} vanilla conjugate gradient method.

In \Figure{apo noise chi2}, we fixed $f_\text{knee}=10$ Hz, and change $f_\text{apo}$.
As we decrease $f_\text{apo}$ relative to $f_\text{knee}$, increasing the amount of low-frequency noise,  the parameterized conjugate gradient method performs better.

Looking at the power spectrum in Figure \ref{power spectrum},
when $f_\text{knee}$ is small, or $f_\text{apo}$ is large, there is not much low-frequency noise.
These situations corrrespond to the left-side plots in Figure \ref{1/f noise chi2} and Figure \ref{apo noise chi2}.
The right-side graphs have significant amount of low frequency noise.
We conclude that the introduction of the slowly-varying $\eta$ parameter improves performance most when there are large low-frequency noise contributions.

We also tried different $1/f$ noise slopes $\alpha$. For $\alpha=2$, the conclusion is the
same as $\alpha=3$. When $\alpha=1$, the low-frequency noise is reduced compared to the cases with steeper slopes, and
the vanilla conjugate gradient method is preferred, except some cases with very large
knee frequency like $f_\text{knee} = 100$ Hz and $f_\text{apo}=0$ which favors the
parameterized method.
In \citet{2018A&A...620A..59P}, the slope $\alpha = 1$ and  the noise power spectrum is flattened at $f_\text{apo} \approx 0.1 f_\text{knee}$. Their knee frequency is the same as their scanning frequency, so is most like our case when $f_\text{knee}=f_\text{scan}=0.1$~Hz.  Their case had little low-frequency noise, and we confirm their specific result that the standard conjugate gradient method converges faster in that case.  In general, however, we find cases with significantly more low-frequency noise benefit from the cooling/parameterized approach.

\section{Conclusions} \label{sec:conclusions}

We analyzed {the} parameterized conjugate gradient map-making method that is inspired by the messenger-field idea of
separating the white noise out of {the} noise covariance matrix.
Then we gave an analytical expression for the series of $\eta$ parameters that govern how quickly the modified covariance adjusts to the correct covariance,
and showed that this method adds only the extra computational cost of re-initializing the conjugate gradient process based on a new $\eta$ parameter.

We tested this method for different noise power {spectra}, both flattened and non-flattened at low frequency.
The results showed that the parameterized method is faster than the traditional conjugate gradient method 
when there is a significant amount of low-frequency noise.
It could be further improved if we could get a more accurate estimation for the change in $\chi^2$ as a function of the $\eta$ parameter, either before iteration or without using time ordered data during iteration.

Also note that we fixed the preconditioner as $M = \Pdagger P$ during our calculation,
this parameterizing process could be applied to any preconditioner and possibly improve performance when 
there is significant amount of low-frequency noise.

This type of analysis for the cooling parameter may also be apply to other areas, like the Wiener filter.
\citet{2018A&A...620A..59P} showed that the messenger field method of \citet{2013A&A...549A.111E} for solving Wiener filter problem  could also be written as a parameterized conjugate gradient algorithm.
It stands to reason that such a system may also benefit from splitting and parameterizing its noise covariance, depending on the noise properties.
(In the Wiener filter, \citet{2017MNRAS.468.1782K} additionally suggests the splitting the signal covariance and combining the uniform parts of the signal and noise.)

The benefits to map-making from a cooled messenger method seem to come from the cooling and not actually from the messenger field that inspired it.  However, the messenger field approach may still have a role in the production and analysis of CMB maps.  In particular, the close connection between the messenger method and Gibbs sampling may allow us to cheaply generate noise realizations of a converged map by generating samples from the map posterior distribution, something that we will continue to explore in future work.

\begin{acknowledgments}
For this work, QBQ and KMH are supported by NSF award 1815887.
\end{acknowledgments}

\appendix
\section{The derivation of $\eta$ parameter series} \label{appendix:eta calculation}

We know that the initial degree of heteroscedasticity $\eta_0 = 0$,
which means the system is homoscedastic (uniform noise) to start.
What would be a good value for the next parameter $\eta_1$?
To simplify notation, we use $\Neta$ to denote the parameterized covariance matrix
$N(\eta) = \tau I +  \eta \bar N$.
For some specific $\eta$ value, the estimated map
$\hatm(\eta) = \PPinv{\inv{\Neta}} \Pdagger \inv{\Neta} \vbd$ minimizes
\begin{align}
\begin{aligned}[b]
\chi^2 (\vbm, \eta)
&= \qty\big(\vbd - P \vbm)^{\dagger} \inv{\Neta} 
    \qty\big(\vbd - P\vbm).
\label{chi2 eta formula}
\end{aligned}
\end{align}
with $\eta$ being fixed.
We restrict to the case that the noise covariance matrix $N$ is diagonal in the frequency domain,
and represent the frequency-domain eigenvalues as $N_f$.

The perturbative scheme works like this.
We start with $\chi^2(\hatm(\eta_0),\eta_0)$ with $\hatm(\eta_0) = \PPinv{} \Pdagger \vbd$ which could be solved directly.
Then we use conjugate gradient method to find $\hatm(\eta_1)$
and
    its corresponding $\chi^2(\hatm(\eta_1),\eta_1)$.
So let us consider $\eta_1 = \eta_0 + \delta\eta = \delta\eta$
such that $\eta_1 = \delta \eta$ is very small quantity, $\delta \eta \ll 1$. (Remember $\eta_0 = 0$.)
Since $\hatm(\eta)$ minimizes $\chi^2(\vbm,\eta)$ with $\eta$ being fixed, we have 
$\pdv{\hatm} \chi^2(\hatm(\eta), \eta) = 0$,
and using the chain rule
\begin{align}
\dv{\eta} \chi^2(\hatm(\eta), \eta) = \pdv{\eta} \chi^2(\hatm(\eta), \eta) 
= - \qty(\vbd - P\hatm(\eta))^{\dagger} \inv{\Neta} \Nbar \inv{\Neta}
    (\vbd - P\hatm(\eta)) \label{d chi2}
\end{align}
Then the fractional decrease of $\chi^2(\hatm(\eta_0),\eta_0)$ from $\eta_0$ to $\eta_1 = \delta \eta$ is
\begin{align}
-\frac{\delta \chi^2(\hatm(\eta_0), \eta_0)}{\chi^2(\hatm(\eta_0), \eta_0)} 
= - \delta \eta \frac{\dv{\eta} \chi^2(\hatm(\eta_0),\eta_0)}{\chi^2(\hatm(\eta_0),\eta_0)}
&= \delta \eta 
\frac{1}{\tau}
\frac{\qty(\vbd - P\hatm(\eta_0))^{\dagger} \Nbar  (\vbd - P\hatm(\eta_0)) }
    {\qty\big(\vbd - P \hatm(\eta_0))^{\dagger} \qty\big(\vbd - P\hatm(\eta_0))}
\label{chi2 fractional decrease}
\end{align}
Here we put a minus sign in front of this expression such that it is 
non-negative, and use $N_{\eta=0} = \tau I$ at the second equality.
We want $\qty|\delta \chi^2(\hatm(\eta_0),\eta_0)| = \chi^2(\hatm(\eta_0),\eta_0)) - \chi^2(\hatm(\eta_1), \eta_1)$
to be large to encourage fast convergence.
Therefore $\chi^2(\hatm(\eta_1), \eta_1)$ is much smaller than $\chi^2(\hatm(\eta_0), \eta_0)$,
or $\chi^2(\hatm(\eta_1), \eta_1) \ll \chi^2(\hatm(\eta_0), \eta_0)$.
Then we would expect
\begin{align}
-\frac{\delta\chi^2(\hatm(0),0)}{\chi^2(\hatm(0),0)}
= 1 - \frac{\chi^2(\hatm(\eta_1),\eta_1)}{\chi^2(\hatm(0),0)}
\approx 1^-
\label{dchi2/chi2 0}
\end{align}
Here we use the notation $1^-$ means the upper bound is close to but strictly smaller than 1.
Now we could use \Eq{chi2 fractional decrease} and let it equal to $1$, then
$\delta\eta=-\chi^2(\hatm(\eta_0), \eta_0)/\dv{\eta}\chi^2(\hatm(\eta_0),\eta_0)$.

Applying this same idea to $\eta_{m+1} = \eta_m + \delta \eta_m$ with $m \geq 1$, we would get
\begin{align}
\delta\eta_m = -\chi^2(\hatm(\eta_m), \eta_m)/\dv{\eta}\chi^2(\hatm(\eta_m),\eta_m).
\label{delta eta update}
\end{align}
As mentioned in the main text, we need to determine the entire series $\qty{\eta_i}$ before conjugate gradient iterations.
We do not have the $\hatm(\eta_m)$ with which to calculate them and need to find another approach.

Let us go back to \Eq{chi2 fractional decrease}.
Since we cannot calculate $\vbd - P\hatm(\eta_m)$ before making the map, 
we treat it as an arbitrary vector, then the least upper bound of \Eq{chi2 fractional decrease} is given by
\begin{align}
-\frac{\delta \chi^2(\hatm(\eta_0), \eta_0)}{\chi^2(\hatm(\eta_0), \eta_0)} 
\leq \frac{\delta \eta} {\tau} \max(\Nbar_f)
\label{dchi2/chi2 upper bound}
\end{align}
where $\max(\Nbar_f)$ is the maximum eigenvalue of $\Nbar$.
We want $ -\frac{\delta \chi^2(\hatm(\eta_0), \eta_0)}{\chi^2(\hatm(\eta_0), \eta_0)}$ to be as large as possible,
but it won't exceed $1$.
If we combine \Eq{dchi2/chi2 0} and \Eq{dchi2/chi2 upper bound},
and choose $\delta \eta$ such that the least upper bound is equal to 1,
to make sure the process would not going too fast.
Thus we have
\begin{equation}
\eta_1 = \delta \eta  = \frac{\tau}{\max(\Nbar_f)} = \frac{\min(N_f)}{\max(N_f) - \min(N_f)}.
\end{equation}
Here $N_f$ and $\Nbar_f$ are the eigenvalues of $N$ and $\Nbar$ in the frequency
domain.
If the condition number of noise covariance matrix
$\kappa(N) = \max(N_f)/\min(N_f) \gg 1$,
then $\eta_1 \approx \inv{\kappa} (N)$.

What about the other parameters $\eta_m$ with $m > 1$?
We use a similar analysis,
letting $\eta_{m+1} = \eta_m + \delta \eta_m$ with a small $\delta\eta_m \ll 1$.
First, let us find the least upper bound
\begin{align}
-\frac{\delta \chi^2(\hatm(\eta_m), \eta_m)}{\chi^2(\hatm(\eta_m), \eta_m)}  
=& \delta\eta_m
\frac{\qty(\vbd - P\hatm(\eta_m))^{\dagger}
    \inv{N_{\eta_m}} \Nbar \inv{N_{\eta_m}}
    (\vbd - P\hatm(\eta_m))
}
{\qty\big(\vbd - P \hatm(\eta_m))^{\dagger}
    \inv{N_{\eta_m}}
    \qty\big(\vbd - P\hatm(\eta_m))
}
\label{dchi2 chi2}
\\
\leq & \delta \eta_m\, \max\qty(\frac{\Nbar_f}{\tau + \eta_m \Nbar_f})
\end{align}
The upper bound in the second line is a little bit tricky.
Both matrix $\Nbar$ and $\inv{N}_{\eta_m}$ 
can be simultaneously diagonalized in frequency space.
For each eigenvector $\vb{e}_f$,
the corresponding eigenvalue of the matrix on the numerator
$\inv{N}_{\eta_m} \Nbar \inv{N}_{\eta_m}$
is
$\lambda_f = \Nbar_f (\tau + \eta_m \Nbar_f)^{-2}$,
and the eigenvalue for the  matrix in the denominator
$\inv{N}_{\eta_m}$
is
$\gamma_f = (\tau + \eta_m \Nbar_f)^{-1}$.
Their eigenvalues are related by
$\lambda_f = [{\Nbar_f}/{(\tau + \eta_m \Nbar_f)}] \gamma_f$.
For any vector $\vb{v} = \sum_f \alpha_f \vb{e}_f$, we have
\begin{equation}
  \frac{\vb{v}^{\dagger} \inv{N}_{\eta_m} \Nbar \inv{N}_{\eta_m} \vb{v}}
{\vb{v}^{\dagger} \inv{N}_{\eta_m} \vb{v}}
= \frac{\sum_f \alpha_f^2 \lambda_f}{\sum_f \alpha_f^2 \gamma_f}
= \frac{\sum_f \alpha_f^2 \gamma_f \Nbar_f/(\tau + \eta_m \Nbar_f)}
{\sum_f \alpha_f^2 \gamma_f}
\leq \max \qty( \frac{\Nbar_f}{\tau + \eta_m \Nbar_f}).
\end{equation}

Again assuming $\chi^2(\hatm(\eta_{m+1}), \eta_{m+1}) \ll \chi^2(\hatm(\eta_m), \eta_m)$,
which we expect it to be satisfied for $ \eta_m \ll 1$.
That is because if $\eta \lesssim 1$, $\chi^2(\hatm(\eta), \eta)$ would close to the minimum $\chi^2$
which means $\chi^2(\hatm(\eta_{m+1}), \eta_{m+1}) \lesssim \chi^2(\hatm(\eta_m), \eta_m)$,
which would violate our assumption.
Luckily, the final result (\Eq{etai rule appendix}) is a geometric series,
only the last few $\eta_m$ values fail to satisfy this condition.
Similarly, we could set the least upper bound equal to 1.
Then we get
\begin{align}
\delta \eta_m 
= \min \qty(\frac{\tau + \eta_m \Nbar_f}{\Nbar_f})
= \eta_m + \frac{\tau }{\max(\Nbar_f)}.
\end{align}
Therefore 
\begin{align}
\eta_{m+1} = \eta_m + \delta\eta_m = 2\eta_m + \frac{\tau }{\max (\Nbar_f)}
\end{align}
If written in the form $\eta_{m+1} + {\tau }/{\max(\Nbar_f)}
= 2 \qty( \eta_m + {\tau}/{\max(\Nbar_f)})$
it's easy to see that for $m \geq 1$,
$\eta_{m} + {\tau }/{\max(\Nbar_f)}$ forms a geometric series
\begin{align}
\eta_m +  \frac{\tau }{\max(\Nbar_f)}
=\qty(\eta_1 + \frac{\tau }{\max(\Nbar_f)}) 2^{m-1}
=\frac{\tau}{\max(\Nbar_f)} 2^m
\end{align}
where we used $\eta_1 = {\tau}/{\max(\bar{N}_f)}$.
Note that $m = 0$ and $\eta_0 = 0$ also satisfy this expression and we've got
final expression for all $\eta_m$
\begin{align}
\eta_m =\min \qty\bigg{1,\; \frac{\tau}{\max(\Nbar_f)} \qty(2^m -1) }
\label{etai rule appendix}
\end{align}
Here we need to truncate the series when $\eta_m > 1$.

In numerical simulations, we find that if we update the $\eta$ parameter based on the more precise \Eq{delta eta update},
there is only marginally improvements over the $\eta$ series given in \Eq{etai rule appendix}
for the $1/f$ noise model, and slight improvements when there is not much low frequency noise.
If we use \Eq{delta eta update},
it ends up with fewer $\eta$ parameters in the series, but the interval between $\eta_i$ and $\eta_{i+1}$ gets larger. 
In our simulation this sometimes causes one more iteration at certain $\eta$ value, so in the end there 
is only slight improvements.
For large data set that need lots of iterations to converge from $\eta_i$ to $\eta_{i+1}$,
where several extra iterations may not be significant, using \Eq{delta eta update} may provide a larger performance
boost.


\bibliography{references.bib}{}
\bibliographystyle{aasjournal}



\end{CJK*}
\end{document}